%% file: vortex.tex
\documentclass[submission,copyright,creativecommons]{eptcs}

\usepackage{hyperref,url}
\usepackage{xspace}
\usepackage{graphicx}
\usepackage{xcolor}
\usepackage{listings}
\usepackage{tcolorbox}
\usepackage{enumitem}
\usepackage{xspace,cite}

\providecommand{\event}{Vortex 2018} % Name of the event you are submitting to
\usepackage{breakurl}             % Not needed if you use pdflatex only.
\usepackage{underscore}           % Only needed if you use pdflatex.
\usepackage{wrapfig}
\usepackage{float}
\floatstyle{plaintop}
\restylefloat{table}

\lstdefinestyle{mystyle}{
language=java,columns=fullflexible,
 		   mathescape=true,%
 		   showstringspaces=false,%
keywordstyle=\bf\ttfamily,
commentstyle=\sl\color{gray},%
basicstyle=\small\ttfamily,
inputencoding=latin1, % i would prefer utf8
extendedchars,xleftmargin=0em
}

\graphicspath{ {fig/} }

\usepackage{tikz,tikzsymbols}
\usetikzlibrary{shapes.geometric} % required for the ellipse shape
\usetikzlibrary{arrows, backgrounds, calc, hobby, positioning,shapes}

\usepackage{pgfplots}
\pgfplotsset{compat=1.11}
\usepgfplotslibrary{fillbetween}
\usetikzlibrary{intersections}

\tikzset{
  center coordinate/.style={
    execute at end picture={
      \path ([rotate around={180:#1}]perpendicular cs: horizontal line through={#1},
                                  vertical line through={(current bounding box.east)})
            ([rotate around={180:#1}]perpendicular cs: horizontal line through={#1},
                                  vertical line through={(current
                                    bounding box.west)});}},
    vertex style/.style={
    draw=#1,
    thick,
    fill=#1!70,
    text=white,
    rectangle,
    rounded corners,
    text width=1.9cm,
    minimum height=0.75cm,
    font=\small,
    outer sep=3pt,
  },
    vertexlight style/.style={
    draw=#1,
    thick,
    fill=#1!20,
    text=black,
    rectangle,
    rounded corners,
    text width=1.9cm,
    minimum height=0.75cm,
    font=\small,
    outer sep=3pt,
  },
    vertextext style/.style={
    draw=#1,
    thick,
    fill=#1!30,
    text=black,
    rectangle,
    rounded corners,
    text width=5cm,
    font=\it\footnotesize,
    outer sep=3pt,
  },
    longvertex style/.style={
    draw=#1,
    thick,
    fill=#1!70,
    text=white,
    rectangle,
    rounded corners,
    text width=2.2cm,
    minimum height=0.75cm,
    font=\small,
    outer sep=3pt,
  },
  diavertex style/.style={
    draw=#1,
    fill=#1!10,
    text=#1,
    ellipse,
    text width=1.6cm,
    minimum height=0.75cm,
    font=\small,
    outer sep=3pt,
    dotted
  },
  text style/.style={
    sloped,
    text=black,
    font=\footnotesize,
    above
  }}

\input{macros}

\newcommand{\java}[1]{{\lstset{language=Java}\lstinline!#1!}}

\newcommand{\FRH}{Fredhopper\xspace}
\newcommand{\frhtool}{Fredhopper Cloud Services\xspace}

\input{java}

\title{Analysis of SLA Compliance in the Cloud:\\
  An Automated, Model-based Approach\thanks{Partially funded by EU project FP7-610582 ENVISAGE:
  Engineering Virtualized Services
  \href{https://www.envisage-project.eu}{\texttt{www.envisage-project.eu}}.}}

\author{
  Frank S. de Boer
  \institute{CWI Amsterdam\\The Netherlands}
  \email{F.S.de.Boer@cwi.nl}
  \and
  Elena Giachino
  \institute{University of Bologna\\Italy}
  \email{elena.giachino@unibo.it}
  \and
  Stijn de Gouw
  \institute{The Open University\\The Netherlands}
  \email{cdegouw@gmail.com}
  \and
  Reiner H\"ahnle
  \institute{Technical University of Darmstadt\\Germany}
  \email{haehnle@cs.tu-darmstadt.de}
  \and
  Einar Broch Johnsen
  \institute{University of Oslo\\Norway}
  \email{einarj@ifi.uio.no}
  \and
  Cosimo Laneve
  \institute{University of Bologna\\Italy}
  \email{cosimo.laneve@unibo.it}
  \and
  Ka I Pun
  \institute{Western Norway University of Applied Sciences\\University of Oslo\\Norway}
  \email{Violet.Ka.I.Pun@hvl.no}
  \and
  Gianluigi Zavattaro
  \institute{University of Bologna\\Italy}
  \email{zavattar@cs.unibo.it} 
}
\def\titlerunning{Analysis of SLA Compliance in the Cloud}
\def\authorrunning{F. de Boer et al.}

\begin{document}
\maketitle

\begin{abstract}
  \textbf{Abstract.} Service Level Agreements (SLA) are commonly used to specify the
  quality attributes between cloud service providers and the
  customers.  A violation of SLAs can result in high penalties.  To
  allow the analysis of SLA compliance before the services are
  deployed, we describe in this paper an approach for SLA-aware
  deployment of services on the cloud, and illustrate its workflow by
  means of a case study.  The approach is based on formal models
  combined with static analysis tools and generated runtime
  monitors. As such, it fits well within a methodology combining
  software development with information technology operations
  (DevOps).
\end{abstract}

\section{Model-Centric Analysis of SLA Compliance}\label{sec:intro}
\input{intro}

\section{What to be Measured? What to be Verified?}\label{sec:metrics}
\input{metrics}

\section{Model-Centric SLA Compliance: The Workflow}\label{sec:flow}
\input{analysis}

\section{ABS: A Modeling Language and Tool Suite\\ for Systems
  Deployed on the Cloud}\label{sec:abs}
\input{abs}

\section{Case Study}
\label{sec:case-study}
\input{monitor}

\section{Related Work}
\label{sec:related-work}
% \begin{itemize}
%\item Paragraph on our Envisage-related work, we can include more
%  details than in the workshop submission since we still have almost 2
%  pages
%  \item check related work, in poarticular from the Behrooz/Stijn ESOCC
%    papers on runtime monitoring of SLA \cite{bezirgiannis17esocc,giachino16fdb,gouw16esocc,boer14taosd,nobakht15esocc,gouw15esocc}
%\item Work of Samir Tata et al \cite{hadded18icac,graietH17facs,mohamed17coop}
%\item Farokhi \cite{farokhi14ccgrfid}
%\end{itemize}
\input{relatedworks}

\section{Conclusion}
\label{sec:conclusion}

This paper describes the analysis SLA compliance for services deployed
on the cloud, by combing the formal models of the SLA and of the cloud
service.  Based on these two formal models, a detailed model-centric,
tool-supported workflow is defined to obtain a configuration of cloud
services in a semi-automated manner.
The basis for our approach is the modeling language ABS that supports
the modeling of deployment decisions on elastic infrastructure, and is
the basis for a scalable monitoring framework for deployed services
based on service metric functions.
Using an industrial case study from Fredhopper Cloud Services, we show
that our model-based approach can help to address the challenging
questions posed in Section~\ref{sec:intro}.
Our specific combination of model-based tools is a good match for a
DevOps methodology.

\bibliographystyle{eptcs}
\bibliography{biblio,reiner} 

\end{document}

%% file: macros.tex
\usepackage{array}
\newcolumntype{L}[1]{>{\raggedright\let\newline\\\arraybackslash\hspace{0pt}}m{#1}}
\newcolumntype{C}[1]{>{\centering\let\newline\\\arraybackslash\hspace{0pt}}m{#1}}
\newcolumntype{R}[1]{>{\raggedleft\let\newline\\\arraybackslash\hspace{0pt}}m{#1}}

%% file: java.tex
\lstdefinelanguage{JavaSpec}{keywords={
     abstract,boolean,break,byte,case,catch,char,class,%
     const,continue,default,do,double,else,extends,false,final,%
     finally,float,for,goto,if,implements,import,instanceof,int,%
     interface,label,long,native,new,null,package,private,protected,%
     public,return,short,static,super,switch,synchronized,this,throw,%
     throws,true,try,void,volatile,while,assert,enum,%
     global, call, view, result, local, specifies, grammar, callee, caller},%
  sensitive=true, comment=[l]{//},
  morecomment=[s]{/*}{*/},
  morestring=[b]"}

\renewcommand{\java}[1]{\lstinline[language=JavaSpec,columns=fullflexible,mathescape=true,keywordstyle=\textbf\sffamily,basicstyle=\normalsize\ttfamily]!#1!}

\lstdefinestyle{javastyle}{
language=JavaSpec,columns=fullflexible,
 		   mathescape=true,%
 		   showstringspaces=false,%
keywordstyle=\textbf\sffamily,
commentstyle=\textsl\sffamily,%
basicstyle=\footnotesize\tttfamily,
inputencoding=latin1, % i would prefer utf8
extendedchars,xleftmargin=2em
}

\lstnewenvironment{javaexamplen}[1][]{
\lstset{style=javastyle,
basicstyle=\footnotesize\ttfamily,
framerule=.5pt,
belowskip=0pt,
backgroundcolor=\color{green!3},
rulecolor=\color{green!50!black},
frame=tblr,
captionpos=b,
xleftmargin=4pt,
xrightmargin=4pt,#1,
mathescape=false
}}
{\vspace{\baselineskip}}

\lstnewenvironment{javaexamplenum}[1][]{
\lstset{style=javastyle,
basicstyle=\footnotesize\ttfamily,
framerule=.5pt,
belowskip=0pt,
backgroundcolor=\color{green!3},
rulecolor=\color{green!50!black},
frame=tblr,
captionpos=b,
xleftmargin=4pt,
xrightmargin=4pt,#1,
mathescape=false,
numbers=left
}}
{\vspace{\baselineskip}}

\lstdefinelanguage{JML}%
       {language=Java,
        morecomment=[s]{/*\ }{*/},
        morecomment=[s]{/**}{*/},
        classoffset=1,
        morekeywords={abrupt_behavior,abrupt_behaviour,
         accessible,accessible_redundantly,also,assert,assert_redundantly,
         assignable,assignable_redundantly,assume,assume_redundantly,
         axiom,behavior,behaviour,breaks,breaks_redundantly,
         callable,callable_redundantly,captures,captures_redundantly,
         choose,choose_if,code,code_bigint_math,code_java_math,
         code_safe_math,constraint,constraint_redundantly,constructor,
         continues,continues_redundantly,decreases,decreases_redundantly,
         decreasing,decreasing_redundantly,diverges,diverges_redundantly,
         duration,duration_redundantly,ensures,ensures_redundantly,
         example,exceptional_behavior,exceptional_behaviour,
         exceptional_example,exsures,exsures_redundantly,extract,field,
         forall,for_example,ghost,helper,hence_by,hence_by_redundantly,
         implies_that,in,in_redundantly,initializer,initially,instance,
         invariant,invariant_redundantly,loop_invariant,
         loop_invariant_redundantly,maintaining,maintaining_redundantly,
         maps,maps_redundantly,measured_by,measured_by_redundantly,method,
         model,model_program,modifiable,modifiable_redundantly,modifies,
         modifies_redundantly,monitored,monitors_for,non_null,
         normal_behavior,normal_behaviour,normal_example,nowarn,
         nullable,nullable_by_default,old,or,post,post_redundantly,
         pre,pre_redundantly,pure,readable,refine,refines,refining,represents,
         represents_redundantly,requires,requires_redundantly,
         returns,returns_redundantly,set,signals,signals_only,
         signals_only_redundantly,signals_redundantly,spec_bigint_math,
         spec_java_math,spec_protected,spec_public,spec_safe_math,
         static_initializer,uninitialized,unreachable,weakly,
         when,when_redundantly,working_space,working_space_redundantly,
         writable
        },
        morekeywords={rep,peer,readonly},
        keywordsprefix=\\,
        otherkeywords={<:,<-,->,..,<==,==>,<==>,<=!=>},
        classoffset=0 % restore default class for keywords
} 

%% file: intro.tex
Every customer wants to be sure about the quality of his
purchases. In the cloud world, this quality assurance includes
guarantees on service \emph{performance}\footnote{Other concerns
  include security, support, data management and data
  protection~\cite{CloudGuidelines2014}.}.

Service Level Agreements (SLAs) are legal documents, signed and agreed
upon by cloud service providers and their customers, which specify the
agreed quality of service.  An SLA violation will result in penalties
and possibly in a loss of money, clients, and credibility.  Even
though the stakes are high, there are only few tools with limited
capabilities available to check the compliance of cloud services with
SLAs.  But why does it seem to be so difficult to provide tool support
for SLA compliance checking and monitoring?

For a start, a number of complex and challenging questions arise: How
to describe service performance?  How many resources, for example,
memory or virtual machines, should be assigned to a particular service
and how should they be configured?  How to react optimally at runtime
to take advantage of the elasticity of the cloud?  How to estimate the
future behavior of a service and adjust the resource configuration
accordingly?

These are challenging issues! It is beyond current technology to
address them in a general way for any given SLA and any given
software.  To develop effective tools for SLA compliance analysis, we
believe it is essential to work at the level of \emph{models}, and
describe and analyze SLAs in a way that is independent of the concrete
technology offered by the cloud service provider. Shifting to the
modeling level increases the level of abstraction, reduces complexity,
and removes dependency on a specific runtime environment.

The importance of models applies to SLAs as well as to software: a
model-centric approach allows us to create a formal representation of
the essential aspects of an SLA. At the same time, software services
deployed on the cloud can be represented as an executable
\emph{service model}, annotated with parametric expressions for their
use of resources.  Combining the two models, i.e.,\ of the SLA and of
the cloud services, makes it now possible to use techniques with a
formal basis, such as static software analysis or monitor
generation. Such tools provide proven guarantees on service
performance, thereby vastly raising the degree of automation.

\medskip

\begin{tcolorbox}[colback=purple!5,colframe=purple!40!black, title=\textbf{Benefits of model-centric, tool-based SLA analysis}]
  \small
  An effective solution to SLA design and compliance must coordinate
  all phases of service provisioning:
\begin{itemize}[leftmargin=10pt]
\item Provide assistance in the configuration of SLA metric bounds and
  provisioning of virtual machines whose resources comply to the services' requirements
%  Provide assistance in the configuration of SLA metric bounds and
%  resource deployment to be compliant with a given service deployed on
%  the cloud
\item Permit automatic monitoring of the service at runtime
\item Enable speedy reaction to an SLA violation and assistance in its
  resolution
\item Support deployment: significant simplification and increased
  automation
\end{itemize}
\end{tcolorbox}

\medskip

In this paper, we present an approach to facilitate SLA-aware
deployment of services on the cloud by combining the formal executable
model of the target system of deployed services and the formal
representation of corresponding SLAs.
We define a detailed workflow that takes advantage of the formal
models by enabling automated tool support at various stages.
With the help of a case study we domonstrate how our approach can be
realized for a real-world cloud service provider.

The paper is organised as follows: Section~\ref{sec:metrics} describes
the cloud service performance metrics that our approach is supposed to
measure and verify, Section~\ref{sec:flow} outlines the workflow of
the model-centric SLA compliance analysis; Section~\ref{sec:abs}
provides a short introduction of the modeling language used in this
approach, Section~\ref{sec:case-study} presents the case study,
Section~\ref{sec:related-work}, and Section~\ref{sec:conclusion}
concludes the paper.

%%% Local Variables:
%%% mode: latex
%%% TeX-master: "vortex"
%%% End:

%% file: metrics.tex
Service performance metrics measure and assess the performance level
of a service, quantitatively and periodically.  Typical metrics fall
into one of the following categories:

\begin{description}
\item[\textbf{Availability}] is the property of a service to be
  accessible and usable on demand. It includes (\emph{i}) the
  \emph{level of uptime}, namely the percentage of time a service is
  up within a defined period; (\emph{ii}) the \emph{percentage of
    successful requests}, namely the number of requests processed
  without an error over the total number of submitted requests;
  (\emph{iii}) the \emph{percentage of timely serviced responses to
    requests}, that is the number of service provisioning requests
  completed within a defined time period over the total number of
  service provisioning requests.

\item[\textbf{Response time}] is the time period between a client
  request event and a service response event. The service metrics used
  to constrain the response time may return either an \emph{average time}
  or a \emph{maximum time}, given a particular form of request.

\item[\textbf{Capacity}] is the maximum amount of a resource used by a
  service. It includes the \emph{service throughput metric}, namely
  the minimum number of requests that will be processed by a service
  in a stated time period.  If there is no extra resources provided,
  the more resources a service requires, the lower the service
  throughput will be.
\end{description}

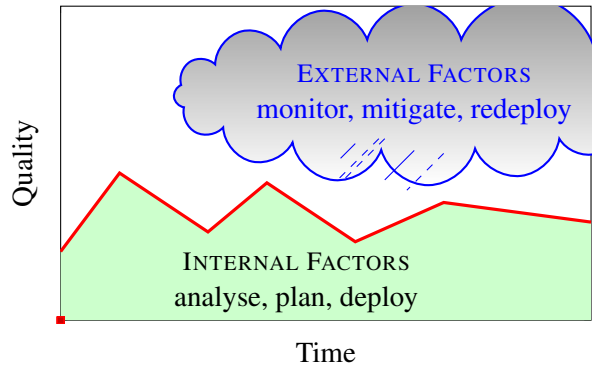
\begin{wrapfigure}{r}{0.5\textwidth}
  \mbox{}
 \bigskip
 \bigskip
 \bigskip
  \input{drawing}
  \caption{Service quality level over time and its influences}
  \label{tab:ServiceQuality}
\end{wrapfigure}

Several factors contribute to the quality level of a service. They can
be classified as \emph{internal}, such as the available resources,
code quality, or the computational complexity, and
\emph{external}. The latter ones are outside the direct control of the
stakeholders and include, for example, network availability or the
number of accesses/requests. The situation is metaphorically
illustrated in Figure~\ref{tab:ServiceQuality}.
Internal factors can principally be controlled (and, if so desired, be
modified) at deployment time with techniques that either directly
verify the code (static analysis) or with the help of an underlying
mathematical model (model checking, simulation, etc.).  Whenever the
service implementation does not comply with the metric, the designer
makes code modifications that eventually lead to compliance.  A
typical example of this is the analysis of the \emph{resource
  capacity} of a service, which measures to which extent a critical
resource is used. For instance, a static analysis technique can
determine an upper bound on the resources needed by a service
\cite{SACO14,GarciaLL17}; if a service is deployed on an insufficient
number of machines, then its response time increases, or it even
becomes unavailable. Thus, internal factors can be expressed and
analyzed inside a model, and integrated into the plans for the initial
deployment of the service on the cloud.

External factors can not be controlled or analyzed in advance, but
they can be supervised by monitoring code that runs independently of
the service implementation.  Monitoring is always needed, as there are
(performance) metrics that are affected by external factors, for
example, hardware failures, which cannot be statically verified.  In
this case, neither the service implementation nor the resource
configuration is at fault. However, a runtime monitor can still be
helpful, for example, when it triggers a dynamic resource
re-allocation that compensates for a faulty component. Thus, external
factors cannot be expressed inside the model and must be monitored on
the deployed service and then mitigated through (static or dynamic)
redeployment of the service.

%%% Local Variables:
%%% mode: latex
%%% TeX-master: "vortex"
%%% End:

%% file: drawing.tex
% \includegraphics[width=12cm]{ReinersDrawing}

% \bigskip

% \begin{table}[h]
%   \centering
%   \begin{tikzpicture}[y=.2cm, x=.7cm,font=\sffamily]
%     \draw[->,name path = lineX] (0,0) -- coordinate (x axis mid)  (16,0);
%     \draw[->] (0,0) -- coordinate (y axis mid) (0,30);
 
%     \draw[Brown,thick,name path = lineA] (0,7)--(2,15)-- (5,9) --
%     (7,14) -- (10,8) -- (13,12) -- (16,10);

%     \node[above=0.3cm,Brown] at (x axis mid) {
%       \begin{tabular}{c} {\sc Internal Factors}\\analyse, plan, deploy
%       \end{tabular}
%     };
     
%     \node[below=0.08cm] at (x axis mid) {Time}; \node[rotate=90,
%     above=0.8cm] at (y axis mid) {Quality};

%     \node[cloud, cloud puffs=15.7, cloud ignores aspect, minimum
%     width=1cm, minimum height=2.5cm, align=center, draw,
%     thick,MidnightBlue] (cloud) at (12,23) {{\sc External
%         Factors}\\monitor, mitigate, redeploy};
 
%   \end{tikzpicture}
  
%   \caption{Service Quality Level}
%   \label{tab:ServiceQuality}
% \end{table}

% \bigskip

% \begin{wrapfigure}{r}{0.5\textwidth} 
%   \mbox{}\bigskip\bigskip\bigskip\bigskip\bigskip
  
%   \hspace{4cm}
  \begin{tikzpicture}[scale=.6]
   \begin{axis}[
     %title=Service Quality Level,
     xmax=18, ymax=32, xmin=0, ymin=0, xlabel= Time, ylabel= Quality,
     xticklabels={,}, yticklabels={,}, xtick={0}, ytick={0}, height=11.5cm, unit vector
     ratio={3 1} ]
     \addplot[red,very thick,name path = A] (0,7)--(2,15)-- (5,9) --
     (7,14) -- (10,8) -- (13,12) -- (18,10); 
     \addplot+[draw=none,name path=B] (0,0)--(18,0); % “fictional” curve
     \addplot+[green!20] fill between[of=A and B];
     % filling
     \node[cloud, cloud puffs=15.7, cloud ignores aspect, minimum
     width=1cm, align=center, draw, thick,
     thick,blue,shading=axis,top color=black!40,bottom color=white] (cloud) at (12,23) {\small\textsc{External Factors}\\monitor, mitigate, redeploy};
     \node at (8,4) {
       \begin{tabular}{c} \textsc{\small Internal Factors}\\analyse, plan,
         deploy
       \end{tabular}
     };
     \draw[blue,dashed] (13,17) -- +(-1.25,-3.75);
     \draw[blue] (12,17.3) -- +(-1,-3);
     \draw[blue,dashed] (11,18.5) -- +(-1.5,-4.5);
     \draw[blue,dashed] (10.8,18.5) -- +(-1.5,-4.5);
     \draw[blue] (10,18) -- +(-.5,-1.5);
   \end{axis}
 \end{tikzpicture} 
%  \caption{Service quality level over time and its influences}
%  \label{tab:ServiceQuality}
% \end{figure}

%%% Local Variables:
%%% mode: latex
%%% TeX-master: "vortex"
%%% End:

%% file: analysis.tex
\begin{figure*}[t!]
\centering
\begin{tikzpicture}[node distance=3.25cm,>=stealth',align=center]
  \node[diavertex style=red] (Sa) {Deployment analysis tools};
  \node[vertex style=blue,above=1.5cm of Sa] (Sm) {Metric\\ functions}
  edge  [->,line width=0.6mm,cyan!60!blue]  (Sa);
  \node[vertexlight style=green,left=4cm of Sm] (Sla) {SLA}
  edge [<-,cyan!60!blue,dashed] node[text style]{compliance violation} (Sa)
  edge [->,cyan!60!blue]  node[text style]{extract}  (Sm);
  \node[vertex style=purple, below=3cm of Sa,yshift=2.5em]
  (Sc) {Service Model}
  edge  [->,line width=0.6mm,cyan!60!blue] node[yshift=.5em] (mm) {} (Sa);
  \node[vertexlight style=yellow, below right=.9cm and 1cm of Sa,yshift=1em] (Rc) 
  {Resource\\ configuration}
  edge [<-,cyan!60!blue,dashed] node[text style]{compliance} node[text style,below]{violation} (Sa);
  \node[vertex style=blue,right=4cm of Sm] (Ma) {Monitor\\ add-on}
  edge [<-,cyan!60!blue] node[text style]{generate} (Sm)
  edge [<-,cyan!60!blue] node[text style]{verify} (Sa);
  \node[text style, above=.1cm of Ma,text=blue,font=\small]{Runtime Techniques};
  \node[vertex style=blue, right=3.8cm of Sa ] (Ra) {Monitoring platform}
  edge [<-,cyan!60!blue] node[text style]{observe/react} (Ma)
  edge [->,cyan!60!blue,dashed] node[text style,xshift=-1em]{(de-)allocate} (Rc.north) 
  edge [->,cyan!60!blue,dashed] node[text style]{feedback} (Sa);
  \node[longvertex style=purple, below=3.2cm of Ra,yshift=2em]
  (Si) {Service\\ implementation}
  edge  [<-,cyan!60!blue] node[pos=.3,text style]{provision} (Rc.south)
  edge  [<->,cyan!60!blue] node[text style]{generate code} node[text style,below]{extract} (Sc)
  edge  [->,line width=0.6mm,cyan!60!blue] (Ra);
\begin{pgfonlayer}{background}
  \draw[blue,fill=blue,dotted,fill opacity=0.1](Ma.north) 
  to[closed,curve through={ (Ma.north west).. %(Ma.west) .. 
    (Ma.south west) 
    % ..($(Ma.south west)!0.5!(Ra.north)$)
    .. (Ra.north     west)%.. (Ra.west) 
    .. (Ra.south west) .. (Ra.south)
    .. (Ra.south east) % .. (Ra.east)
    .. (Ra.north east) %.. ($(Ra.north)!0.35!(Ma.south east)$) 
    .. (Ma.south east) %.. (Ma.east)
    ..(Ma.north east)}](Ma.north);
\end{pgfonlayer}
\draw[<-,line width=0.6mm,cyan!60!blue](Sa.south) to[out=-30,in=180] (Rc.west) ;
\draw[->,cyan!60!blue,dashed](Sa.south) to[out=-75,in=155] node[text style,sloped,pos=.3]{test cases} node[text style,sloped,pos=.6]{compliance} node[text style,sloped,pos=.83]{violation} (Si.west) ;
\draw[->,cyan!60!blue,dashed](Sa.north west) to[out=90,in=230] node[text style,sloped,above]{compliance} node[text style,sloped,below]{violation} (Sm) ;
\end{tikzpicture}
\caption[workflow of service configuration and deployment]{workflow of service configuration and deployment}
\label{fig:flow}
\end{figure*}
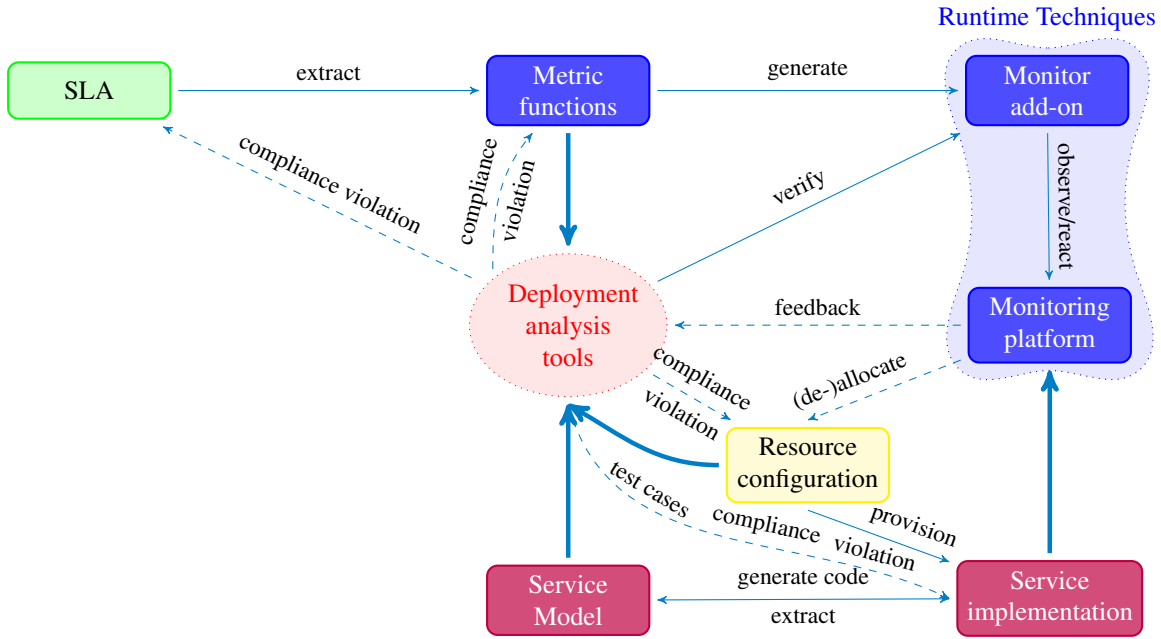

In Section~\ref{sec:intro}, we argued that a model-centric
representation of SLAs and services constitutes the basis for advanced
tool support for cloud service configuration and deployment. In
Section~\ref{sec:metrics}, we explained how the service quality is
influenced by internal factors, to be addressed by compliance checking
of service implementations against SLAs; and by external factors, to
be mitigated with the help of runtime monitors.  In
Figure~\ref{fig:flow}, we illustrate a workflow that realizes
model-centric configuration of cloud services to optimize SLA
compliance under internal and external factors.  The workflow is
divided into \emph{three} phases, namely, Negotiation, Observation and
Reaction.  

Static (deployment analysis) techniques play an important role in
generating initial metrics and monitors that are used in run-time
techniques. \emph{Feedback loops}---represented by dashed arrows---to a
previous phase of the analysis, represent modifications to the system
that ensure, for example, continued compliance after external changes.
Thick blue arrows indicate tool inputs.

\paragraph{Negotiation phase.}
This phase includes everything that might happen before signing a
prospective SLA.  At this stage the SLA metrics are set, so that the
\emph{service model}\footnote{Phrases typeset in italics correspond to
  the artifacts in rectangle boxes in Figure~\ref{fig:flow}.} can be
verified against them.  The \emph{SLA} (top-left corner) is written in
a machine-readable standard format (ISO 19086-2).  Quality-of-service
upper bounds, expressed in terms of \emph{metric functions} over
possible service measurements, are extracted from it. An initial
\emph{resource configuration} is defined over the types of resources
that are allocated for the service (such as CPUs, memory, bandwidth,
etc.). It can be specified manually, or it can be computed
automatically by a solver that returns an optimal distribution of
resources to service instances, given the knowledge of the initial
instances to be deployed, their required computing resources and the
resource costs \cite{zephyrus2}.  At the same time, an executable
\emph{service model} is extracted from the components of the actual
\emph{service implementation}.  The system is provisioned and deployed
using the initial resource configuration.
% (If the actual service exists in advance, it is configured rather
% than generated from scratch.)

A suite of \emph{deployment analysis tools} now takes the three inputs
(Metric functions, Service model, and Resource configuration) and
produces responses as output to form a feasibility assessment.  The
tools can verify properties such as: upper bounds of the resource
consumption (bandwidth, virtual machines, memory allocation, CPU
processing cycles), liveness (deadlock-freedom) and safety
(functionality).  If the tools report that a service model violates an
SLA constraint, then either the constraint can be relaxed or the
resource allocation can be suitably enlarged during the negotiation
phase (with a possible charge for the client). The tools can also
produce test cases that can be used to validate that the service model
captures the implementation.

Thus, the feedback provided by the deployment analysis in terms of
compliance violations guides the negotiation phase by discarding
resource configurations that hinder the ability of the service to meet
the SLA. Feedback may also be used to select a better metric bound,
given the available resources. A third feedback loop may connect back
to the program and allow changes in the code to be applied, so as to
better adapt to the given SLA and available resources.

Once the configuration is approved by the deployment analysis
tools---guaranteeing that, \emph{in the absence of external factors},
the service implementation and the resource configuration comply to
the SLA---the next phase can start.

\paragraph{Observation phase.}
The SLA is now signed and the service implementation is up and running.
%At this stage,
Factors under external control, such as the network
infrastructure, may affect the behavior of the service in ways that
could not be predicted statically.  To supervise the service
metrics we use a monitoring system, namely code external to the
service that continuously monitors its execution and uses self-healing
to repair resource failures or mitigate SLA violations.

The code of the \emph{monitor add-ons} is automatically generated (or
configured), starting from the specific metric functions they are
intended to monitor.  Static techniques may be used at this stage for
proving the correctness of the generated code, i.e., that the monitors
are observing the right property.  Moreover, static techniques may be
performed again at runtime, periodically, on the service model, to
estimate the future behavior of the service in a next time
window. Feedback from the monitoring system can significantly augment
the precision of the analysis.

\paragraph{Reaction phase.}
System monitoring lets the service provider report violations of the
agreed SLA via a \emph{monitoring platform}.  However, the ultimate
goal for a provider is to \emph{dynamically adapt} the resource
configuration so that SLA violations remain under a penalty threshold
while minimizing the cost of the running system.  This can be achieved
by adding appropriate resources to the service (e.g., scaling up the
number and/or size of the virtual machines).

The observation phase takes measurements on services. If an SLA
mismatch is observed by the monitoring platform, in the reaction
phase, the number of allocated resources is increased or decreased
accordingly. As was done for the initial configuration, also in this
phase the modification of the resources assigned to objects can be
done either manually or automatically. A solver computes which new
resources are required and how new service instances should be
distributed on these resources, or how
old objects and resources that are no longer necessary should be
un-deployed, given the knowledge of the current resource configuration
and the new requirements indicated by the monitoring framework. Fully
automatic dynamic elasticity can be obtained thanks to the combined
use of the monitoring framework and the external deployment solver
\cite{gouw16esocc}.

%%% Local Variables:
%%% mode: latex
%%% TeX-master: "vortex"
%%% End:

%% file: abs.tex
ABS \cite{ABSFMCO10} is a modeling language which can be used to
realize model-centric analysis of SLA compliance according to the
workflow outlined in Section~\ref{sec:flow}. We briefly summarize the
main relevant features of ABS
% The box on the facing column gives a concise overview of ABS For
% more information,
(see \href{http://www.abs-models.org}{\texttt{www.abs-models.org}}).

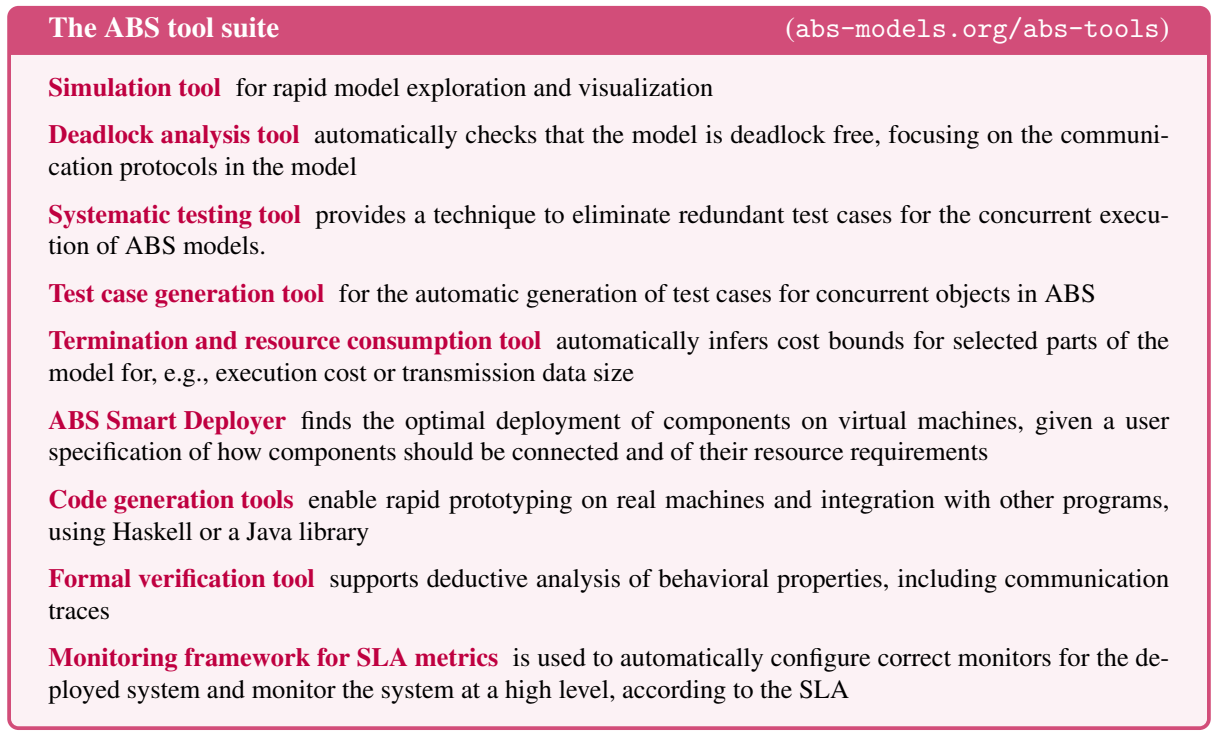
\begin{figure}[t]
  \centering
  \input{abs-tool}
  % \vspace{-4mm}
  \caption{ABS tool suite}
  \label{fig:abs.tool}
\end{figure}

ABS is a language for \textbf{A}bstract \textbf{B}ehavioral
\textbf{S}pecification, which was designed for analyzability.
It combines implementation-level specifications with verifiability,
high-level design with executability, and formal semantics with
practical usability.
ABS is a concurrent, object-oriented modeling language built around a
simple functional language with user-defined algebraic
datatypes. Models are easy to understand and written in a familiar,
Java-like syntax.
In addition, ABS enables replaying a real-world log in the
corresponding executable model through a so-called Model
API~\cite{schlatte18beasts}.
It also explicitly supports the modeling of resource consumption on
virtual machine instances \cite{johnsen15jlamp}.
Thus, the language allows analysis of \emph{deployment decisions}, including
a configurable model of cloud provisioning \cite{HahnleJ15},
and has been used for industrial case studies~\cite{ABHJSTW13}.
Both the \emph{resource requirements} and \emph{timing properties} of
models can be expressed and analyzed, which makes it easy to compare
deployment decisions at the level of models~\cite{WongAMPSS12}
by means of a large portfolio of analysis and
deployment tools (see Figure~\ref{fig:abs.tool}).

% ABS is a modeling language that. It explicitly supports the modeling of resource
% consumption on virtual machine instances \cite{johnsen15jlamp} and has
% been used for industrial case studies \cite{ABHJSTW13}. One of its
% strengths is the availability of a large portfolio of analysis and
% deployment tools \cite{WongAMPSS12}, see the box on the following
% page.

% \input{abs-box}

%%% Local Variables:
%%% mode: latex
%%% TeX-master: "vortex"
%%% End:

%% file: abs-tool.tex
\begin{tcolorbox}[colback=purple!5,colframe=purple!70, title=\textbf{The
  ABS tool suite}~\hfill(\href{http://abs-models.org/abs-tools/}{\texttt{abs-models.org/abs-tools}})
]
\small
  \begin{description}[leftmargin=0cm]
  \item[\textcolor{purple}{Simulation tool}] for rapid model
    exploration and visualization
  \item[\textcolor{purple}{Deadlock analysis tool}] automatically
    checks that the model is deadlock free, focusing on the
    communication protocols in the model
  \item[\textcolor{purple}{Systematic testing tool}] provides a
    technique to eliminate redundant test cases for the concurrent
    execution of ABS models.
  \item[\textcolor{purple}{Test case generation tool}] for the
    automatic generation of test cases  for concurrent objects in ABS 
  \item[\textcolor{purple}{Termination and resource consumption tool}]
    automatically infers cost bounds for selected parts of the model
    for, e.g., execution cost or transmission data size
  \item[\textcolor{purple}{ABS Smart Deployer}] finds the optimal
    deployment of components on virtual machines, given a user
    specification of how components should be connected and of their
    resource requirements
  \item[\textcolor{purple}{Code generation tools}]  enable rapid
    prototyping on real machines and integration with other programs, using Haskell or a Java library
  \item[\textcolor{purple}{Formal verification tool}] supports
    deductive analysis of behavioral properties, including
    communication traces
  \item[\textcolor{purple}{Monitoring framework for SLA metrics}]
    is used to automatically configure correct monitors for the
    deployed system and monitor the system at a high level, according
    to the SLA
  \end{description}
\end{tcolorbox}

%%% Local Variables:
%%% mode: latex
%%% TeX-master: "vortex"
%%% End:

%% file: monitor.tex
The company \FRH provided the \frhtool\footnote{\FRH was recently
  acquired and integrated into the ATTRAQT Group plc, see
  \href{http://www.fredhopper.com}{www.fredhopper.com}} to offer
search on a large product database to e-Commerce companies as services
(SaaS) over the cloud computing infrastructure (IaaS).  At the time of
the case study, \frhtool powered over 350 global retailers with more
than $16$ billion \$ in online sales per year. A customer (service
consumer) of \FRH is a web shop, and an end user is a visitor to the
web shop.

%The company \FRH\footnote{\url{http://www.fredhopper.com}}
%provides \frhtool to offer search and targeting facilities on a large
%product database to e-Commerce companies as services (SaaS) over the
%cloud computing infrastructure (IaaS). \frhtool drives over 350 global
%retailers with more than $16$ billion in online sales every year. A
%customer (service consumer) of \FRH is a web shop, and an end user is
%a visitor to the web shop.

\begin{figure}[t]
  \centering
  \includegraphics[width=0.9\linewidth]{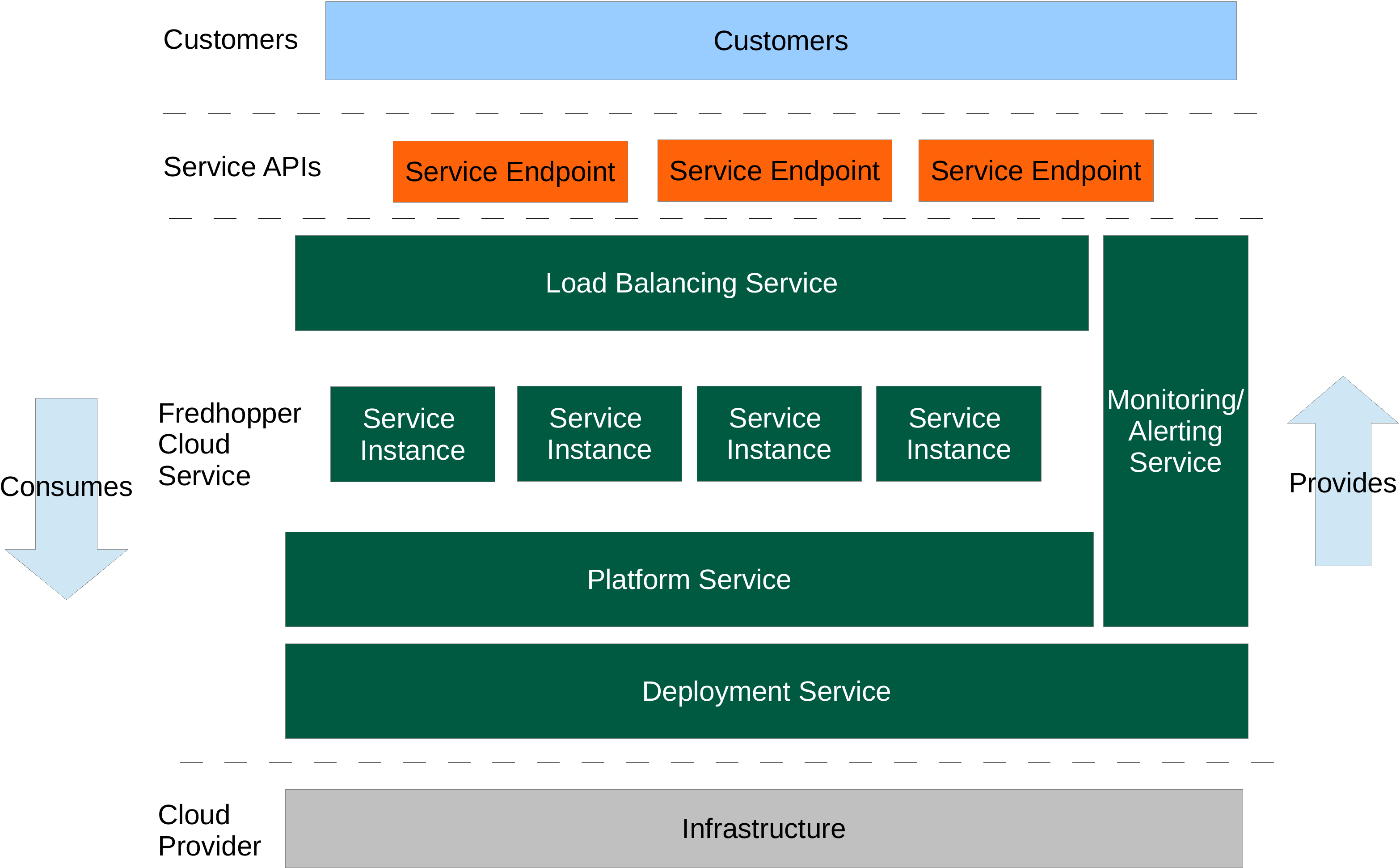}
  \caption{\label{frh-diagram} The architecture of the \frhtool}
\end{figure}

%The services offered by \FRH are exposed at load balanced endpoints.
%In practice, these services are implemented to be RESTful
%and accept connections over HTTP.
Software services offered by \FRH are RESTful and deployed as 
\emph{service instances} that accept connections over HTTP.
Each instance offers the same service and is exposed via Load Balancer
endpoints that distribute requests using a round-robin strategy
over the service instances.
Figure~\ref{frh-diagram} shows a block diagram of the \frhtool.

\begin{figure}[t!]
\centering
\includegraphics[width=0.9\linewidth]{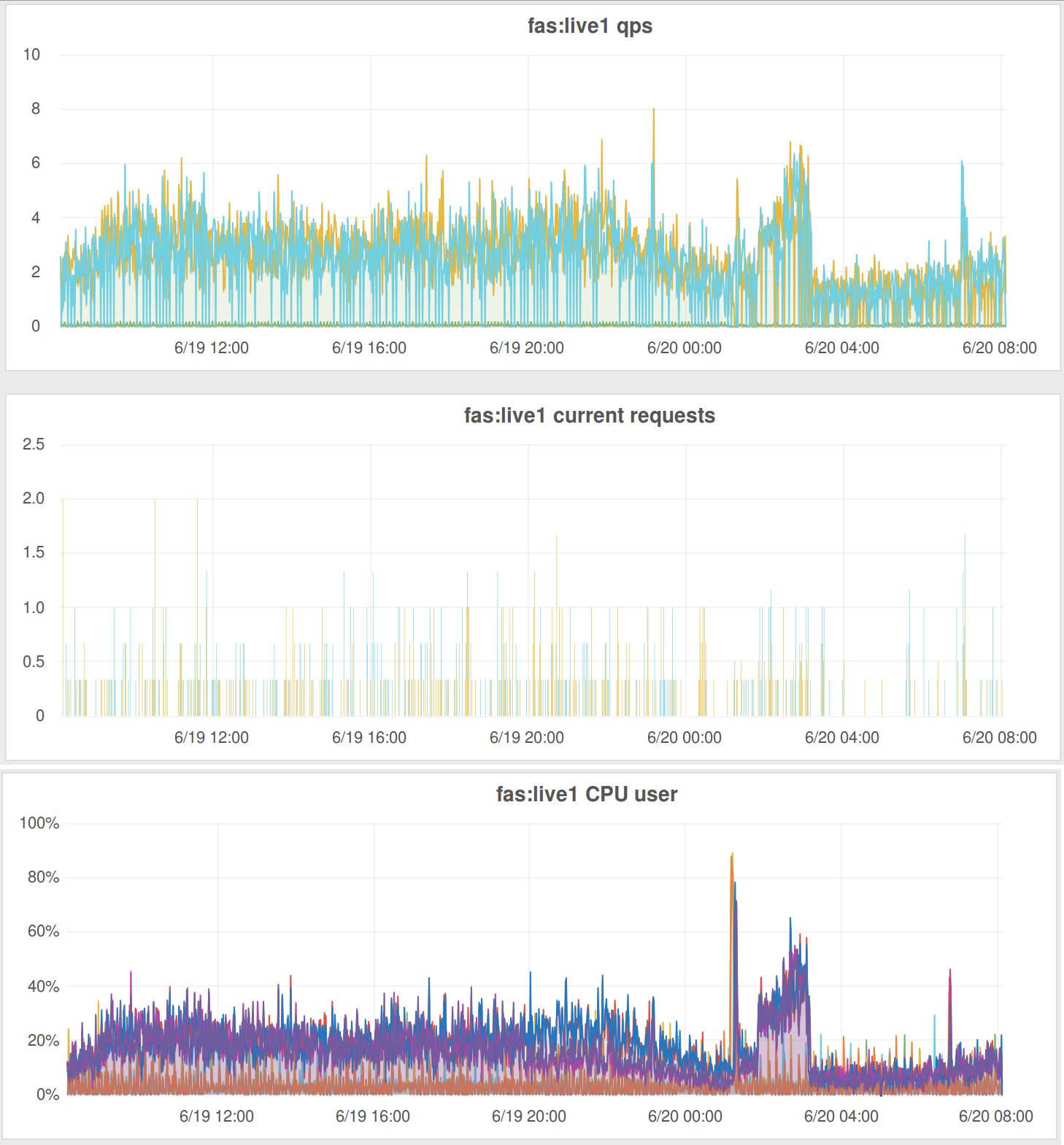}
\caption{\label{fig-metrics}Visualization of metrics}
\end{figure}

The number of requests can vary greatly over time, and typically
depend on several factors.  For instance, the time of the day
where most of the end users are located plays an important
role. Typical lows in demand are observed between 2 am and 5 am.
Figure~\ref{fig-metrics} shows a visualization of monitored data in
Grafana, the visualization framework used by ABS. The top graph shows
the number of query's completed per second (qps), the middle graph
(current requests) shows the number of concurrently served requests
averaged over all service instances of the customer, and the downmost
graph visualizes the average CPU usage over time.

\paragraph{SLA.}
Peaks in demand of Fredhopper Cloud Services typically occur during promotions
of the web-shop or around Christmas.  To ensure a high quality of service,
web shops negotiate an aggressive Service Level Agreement (SLA) with
\FRH.  QoS attributes of interest include query latency
(\emph{response time}) and throughput (\emph{queries per second}).
The SLA negotiated with a customer could express, e.g.,
\emph{service degradation} requirements as follows:

%\begin{quote}\emph{``Services must maintain 100 queries per second
%    with less than 200 milliseconds of response time over 99.5\% of
%    the service up-time.
%    and 99.9\% with less than 500 milliseconds.''}
%  \end{quote}
\vspace{2mm}
\begin{tabular}[h!]{C{1.5cm}C{10cm}R{2.6cm}}
  &\emph{``Services must respond to queries
    in less than 200 milliseconds over 99.5\% of the service up-time.''}&(a)\label{sla.ex}
\end{tabular}
\vspace{2mm}

% \begin{quote}%
%   \emph{``Services must respond to queries
%     in less than 200 milliseconds over 99.5\% of the service up-time.''}
% \end{quote}
\noindent
An SLA specifies properties of service metric functions.  
For the example SLA, the service metric function is defined as the
percentage of client requests which are processed in a ``slow''
manner, i.e., the percentage of queries slower than 200 milliseconds.

\paragraph{Formalizing the service metric function.}

In ABS we formalize a service metric function using an attribute
grammar as a partial mapping of \emph{ traces of events} to values.
The events represent client interactions with an endpoint of an
exposed service API.  The mapping can be partial to detect and exclude
illegal orderings of service invocations.  The values correspond to
different levels of the provided quality of service (QoS).  The
definition of the attributes is given for each production in the form
of ABS code. This ensures that the attribute values are computable and
sufficiently expressive (ABS is Turing complete) to capture general
metrics.  To establish whether a trace of service events is legal, and
if so, what QoS level it should give rise to, the event trace is
parsed according to the grammar. As such, grammars are a user-friendly
formalism and are particularly well-suited for the specification of
both data- and protocol-oriented properties of event traces.  All
regular grammars (with attributes) are currently supported. In such
grammars, each production contains at most one non-terminal which must
occur as the very last symbol in the production. Regular grammars can
be parsed efficiently and incrementally, i.e. there is no need to
re-parse the entire event trace.\footnote{No such method is known for
  general context-free grammars, and it is unlikely to exist as this
  would give a procedure to parse them in linear time (in the size of
  the trace/sequence).}

To formalize our service degradation metric, we identify the
processing of a client request sent to an endpoint of the
exposed query service API by an event

\begin{center}
\java{invoke(Time t, Rat procTime)} 
\end{center}

This event indicates that the request has been issued at time \java{t}
and that it has processing time \java{procTime}.  In our
formalization, a \emph{service view} identifies all the events that
are relevant for a particular service metric and associates a name to
each such event. These names will be used as grammar terminals.
Since there can be many SLAs in a managed cloud service, and each SLA
may concern a different subset of events from the service API(s),
service views allow users to select only those events relevant
for that SLA.
For simplicity, we assume that we treat all requests in the same way.  
A view that simply identifies the invoke event as
the only relevant event and associates the name ``query'' with this
event, is expressed as follows:

\begin{center}
  \begin{minipage}[h!]{0.8\linewidth}
    \begin{lstlisting}[style=mystyle]
      view Degradation { invoke(Time t, Rat procTime) query }
    \end{lstlisting}
  \end{minipage}
\end{center}

Figure~\ref{fig:degradation} contains a grammar\footnote{With the
  usual semantics of CFG's, this grammar generates the empty language:
  no words (sequence of query's) of finite length are derivable. An
  epsilon production would have to be included. As a convenience to
  the user, we allow to omit epsilon productions by using the
  \emph{prefix-closure} of the given grammar. For the given grammar,
  this is all finite sequences of query events.} that computes as the
main metric the percentage of slow queries ``degradation''. The string
``fas.200" gives the name of the metric.  The parameters of the invoke
event, e.g., ``procTime'', are directly referred to in the grammar by
their name and are used to compute the ``degradation'' percentage.
The grammar further makes use of the auxiliary concepts ``cnt'', the
total number of queries, and ``slowCnt'', the total number of ``slow
queries''.

\begin{figure}[th]
  \begin{center}
    \begin{minipage}[h!]{0.6\linewidth}
      \begin{lstlisting}[style=mystyle]
Pair<String, Rat> degradation = Pair("fas.200", 0); 
Int cnt     = 0;
Int slowCnt = 0;

S ::= query
      { cnt = cnt + 1;
        slowCnt = slowCnt +
                   case (procTime > 200) {True   =>  1;
                                             False => 0;};
        degradation = Pair("fas.200", 100 * slowCnt / cnt);
      }
      S
\end{lstlisting}
\end{minipage}
\end{center}
\vspace{-10pt}
\caption{Grammar for Service Degradation}
\label{fig:degradation}
\vspace{-5mm}
\end{figure}

\paragraph{The resource-aware service.}

We create an abstract \emph{service model} in ABS of the various
services shown in Figure~\ref{frh-diagram}.  A detailed
model of the services is not necessary to exploit the ABS tool-set,
it suffices to create a course-grained model that captures the Service APIs
with stub implementations, as we shall see.
The service model can be refined with more detailed implementations
whenever necessary to allow more detailed analyses.
By way of example we show
the model of a Query Service (Figure \ref{fig:ServiceIn}) and the
Load Balancing Service (Figure \ref{fig:LoadB}).  The load balancer
distributes requests by means of a round robin policy and forwards
them to query service instances. (Here, \java{current} is the number of
service instances available in the current round and \java{services}
the instances available in the next round, which may change
dynamically depending on the scaling policy.) The actual service instances process
the requests and return a response, e.g., a list of products that match
the query in the case of Fredhopper Cloud Services.
The given ABS model abstracts
from a detailed implementation and focuses on execution cost by means of
the statement \java{[Cost: cost] log = log $+$ 1}.  The annotation
\java{[Cost: cost]} is a measure of the estimated number of
instructions. An initial value for it can be obtained by using the
SACO tool \cite{SACO14} for cost analysis of models in the ABS tool
suite \cite{WongAMPSS12}, or by averaging execution times from real-world
client logs produced from existing code.

\begin{figure}[ht]
  \begin{center}
    \begin{minipage}[h!]{0.6\linewidth}
      \begin{lstlisting}[style=mystyle]
class QueryServiceImpl (...) implements QueryService {
  ...
  Response invoke (Request request) {
    assert state == RUNNING;
    Int     cost = cost(request);
    Int     time = currentms();
    [Cost: cost] log = log + 1;
    time         = currentms() - time;
    latency      = max(latency, time); 
    return success();
  }
}
\end{lstlisting}
    \end{minipage}
  \end{center}
\vspace{-10pt}
\caption{\label{fig:ServiceIn} Query Service}
\vspace{-5mm}
\end{figure}

\begin{figure}[ht]
  \begin{center}
    \begin{minipage}[h!]{0.6\linewidth}
\begin{lstlisting}[style=mystyle]
class LoadBalancerEndPointImpl
implements LoadBalancerEndPoint {
  Int log     = 0;
  State state = STOP;
  List<QueryService> services = Nil;
  List<QueryService> current  = Nil;
  ...
  Response invoke (Request request) {
    log          = log + 1;
    assert state == RUNNING;
    if (current == Nil) { current = services; }
    EndPoint p   = head(current);
    current      = tail(current);
    return await p!invoke(request);
  }
  ...
}
\end{lstlisting}      
    \end{minipage}
  \end{center}
\vspace{-10pt}
\caption{Load Balancing Service Endpoint}
\label{fig:LoadB}
\vspace{-5mm}
\end{figure}

\paragraph{Negotiation phase.}  Before we can accept a proposed SLA,
we need to determine whether we can meet it with appropriate expense
by deploying a number of \java{QueryServiceImpl} instances.  We
assume a setting where \java{QueryServiceImpl} instances run on
virtual machines with an allocated \emph{capacity} of \java{K}
execution resources (CPU execution capacity, also called ECU).

Static analysis with SACO \cite{SACO14} yields \java{cost/K} as
the total time required by the \java{invoke} method to reply to a
single query.  Therefore, we obtain \java{(cost/K)$\leq$ 0.2} as
a first bound from the SLA~(a) on page~\ref{sla.ex}.  In order to meet
the \emph{service degradation} requirement expressed in the SLA~(a),
we need to determine the minimum number of resources in a
configuration that complies with the SLA. For simplicity, we here
assume a uniform arrival time for the requests, ignore the overhead of
load balancing and distribution, and let~\java{n} be the number
of machines with~\java{k} execution resources that we need. In
this case, we know that
\java{(cost/(n$\times$k))$\leq$ 0.2}, and we obtain
\java{(5$\times$cost/k)$\leq$ n}.  For more complex scenarios
(especially involving sub-services and synchronization), the ABS tool
suite \cite{WongAMPSS12} comes in handy to help calculating the
required number of machines.

This ignores the actual arrival time of requests as well as any
\emph{external} factors (see Figure~\ref{tab:ServiceQuality}) which
may disrupt service execution.  To ensure compliance to the service
metrics under non-ideal conditions, we use a \emph{monitoring
  platform}, external to the service, that continuously observes it.

\paragraph{The observation phase.}
The observation phase in our framework \cite{BG14} consists of
computing the value of the service metric function as specified by the
grammar in Figure~\ref{fig:degradation} from a given event trace.
This involves parsing the event trace according to the grammar.  From
the grammar we automatically synthesize an ABS implementation of the
corresponding parser.  The use of grammars allows us to build on
well-established and widely known parsing technology with optimal
performance. Observations can also come from external systems 
which publish events to the model using an API over HTTP.

\begin{figure}[t]
\centering
\includegraphics[width=0.9\linewidth]{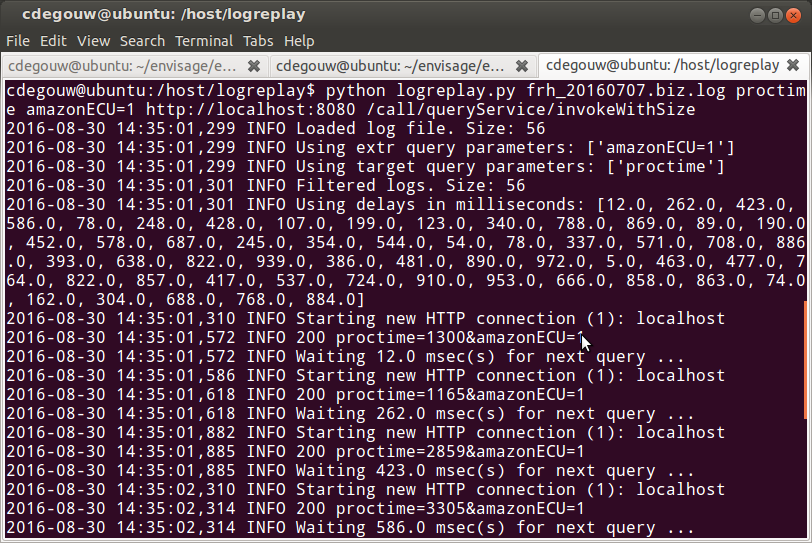}
\caption{\label{fig:logreplay}Log replay}
 \end{figure}

Given our \emph{service model} in ABS, we can now replay a
\emph{real-world log} using this API, which generates corresponding
invoke events for the model according to the specified timings in the
logfile (see Figure~\ref{fig:logreplay}).  The resulting trace of
invoke events is then parsed according to the grammar in order to
compute the ``degradation'' service metric.

\paragraph{Reaction phase.}
Figure~\ref{fig:Monitor} shows a monitor corresponding to the grammar
in Figure~\ref{fig:degradation} for service degradation.  Here
\java{metricHist} contains the time-stamped history of metric values
which is provided by the general ABS monitoring framework.  The
monitoring framework further integrates a powerful tool (the ABS Smart
Deployer~\cite{gouw16esocc}) for the automated deployment of new service
instances, based on high-level requirements of deployment
configurations. A solver synthesizes a provisioning script executable
in ABS that implements \java{DeployerIF} with appropriate scaling
actions, such as allocating new virtual machines, and configuring and
deploying additional service instances on these machines.  This
approach guarantees that the scaling actions preserve the deployment
requirements.

\begin{figure}[ht]
  \begin{center}
    \begin{minipage}[h!]{0.43\linewidth}
\begin{lstlisting}[style=mystyle]
Unit monitor (DeployerIF deployer) {
  Rat degradation = head(metricHist);
  if (degradation > 5/1000) {
      deployer.scaleUp();
  } else if (degradation < 1/1000) {
      deployer.scaleDown();
  }
}
\end{lstlisting}      
    \end{minipage}
  \end{center}
\vspace{-13pt}
\caption{Monitor for Service Degradation}
\label{fig:Monitor}
\vspace{-5mm}
\end{figure}

The above ABS monitor reacts to the service degradation metrics (cf.\
Figure~\ref{fig:degradation}) by
asking the deployer to scale up or down the service instances.
For instance, if degradation is larger than \java{5/1000} (cf.\
SLA~(a)), the method \java{scaleup} is invoked to get more
service instances; and if degradation is less than
\java{1/1000}, the method \java{scaleDown} is invoked to
reduce service instances.
Monitoring
can be expensive; we must ensure that the monitoring does not degrade
performance below the level stipulated in the SLA. Static analysis and
simulation of the ABS model \emph{together} with the monitor allows to
analyze how the monitor effects the SLA \emph{before} the system is
deployed. ABS allows monitors to be deployed asynchronously and
decoupled.

%%% Local Variables:
%%% mode: latex
%%% TeX-master: "vortex"
%%% End:

%% file: relatedworks.tex
The methodology presented in this paper has been devised in the
context of the EU project Envisage to provide efficient development of
SLA-aware and scalable services, supported by highly automated
analysis tools using formal methods.
% SLA-aware services are able to control their own resource management
% and renegotiate SLA across the heterogeneous computing landscape of
% cloud systems.

While there are several proposals for formalizing
SLAs~\cite{KellerL03,LamannaS03}, there is no study on how such SLAs
can be used to both verify and monitor the service and upgrade it as
necessary. In this respect, to the best of our knowledge, our
technique that uses both static analysis and run-time analysis is
original.
% (A method to translate the SLA specification into an operational
% monitoring specification is defined in~\cite{MahbubS11}; this method
% is used by the EU Project SLA@SOI.)
Below we report the main related work on analysis, deployment and
runtime monitoring of systems, focussing on work which is relevant in
the context of cloud systems.

Static analysis estimates computational complexity (e.g.~time) or
resource usage of a given program and provides guarantees that the
program will not exceed the inferred amount of
resources~\cite{GulwaniM09,HoffmannH10}. Typically, such analyses
apply to traditional sequential applications and, in order to use the
above techniques in our context, we had to study the non-trivial
extension to concurrent active object
systems~\cite{SACO14,AlbertCR16,GarciaLL17,GiachinoJLP15}.  The reader
is pointed to the related work sections of these papers for a thorough
comparison with the literature.

The problem of translating customer expectations into metrics to be
measured at runtime is addressed, e.g., by WSLA~\cite{KellerL03} which
introduces a framework to define and break down customer agreements
into a technical description of SLAs and terms to be
monitored. In~\cite{MahbubS11}, a method is proposed to translate the
specification of SLA into a technical domain directed in SLA@SOI EU
project. In the same project~\cite{ComuzziK09} defines terms such as
availability, accessibility and throughput as notions of SLA.
% , however, the formal semantics and properties of the notions are
% not investigated.
In~\cite{ChenI07} the authors describe how they introduce a function
to decompose SLA terms into measurable factors and how to profile
them.  The problem of actually monitoring metrics and react to such
observations, has been addressed e.g.~in MONINA~\cite{ChristianW14},
which is a DSL with a monitoring architecture which supports certain
mathematical optimization techniques, and offers two pre-defined
parameters that can be used in monitoring to adapt the system: cost
and capacity. Also a prototype implementation is available. Hogben and
Pannetrat \cite{HogbenP13} examine the challenges of defining and
measuring availability to support real-world service comparison and
dispute resolution through SLAs.
% They show how two examples of real-world SLAs would lead one service
% provider to report 0\% availability while another would report 100\%
% for the same system state history but using a different period of
% time.

In the context of cloud computing, the problem of automating
application deployment has attracted a lot of attention and many
system management tools
exist~\cite{kanies-puppet-login,mcollective,chef,ansible}.  These
support the specification of deployment plans but not automatic
distribution of the available computing resources to the services to
be deployed. For these reasons, these tools can be seen as deployment
engines to concretely execute deployment plans.  Our approach to
automatic optimal deployment is inspired by the Aeolus component model
\cite{infCom14,concur15} and the Zephyrus configuration optimizer
\cite{zephyrus2}.  The Aeolus model paved the way to formally reason
about deployment and reconfiguration while Zephyrus is a configuration
tool grounded on the Aeolus model.  The use of Aeolus/Zephyrus
guarantees that the synthesized deployment plans are optimal, in the
sense that the total cost of the acquired resources is minimal.
Optimality is guaranteed by taking into account the entire application
architecture. This contrasts with the current auto-scaling
technologies~\cite{swarm,mesos,cloudwatch,Hightower}, supporting the
automatic increase or decrease of the number of instances of a
service, when some conditions (e.g., CPU average load greater than
80\%) are met.  As auto-scaling monitors and scales single specific
services, only local properties can be guaranteed.

%%% Local Variables:
%%% mode: latex
%%% TeX-master: "vortex"
%%% End: